# Preparing Future-Ready Learners: K12 Skills Shift and GenAI EdTech Innovation Direction

Xin Miao[a], Pawan Mishra[a,b]

[a]Alef Education Consultancy L.L.C., Al Muntazah, Ministries Complex, P.O. Box 769719, Abu Dhabi, United Arab Emirates
[b]IIIT-Allahabad, Prayagraj, Uttar Pradesh, India

Since Generative AI came out it has quickly embedded itself in our social fabric, triggering lots of discussions, predictions, and efforts from research, industry, government and capital market to experiment and embrace the technology. The question for the global K12 education is, what and how should our children learn in this fast changing world to be prepared for the changing labor market and live a happy and balanced life? Three key aspects will be discussed: 1) Skills; 2) Evaluation of Learning; 3) Strategic GenAI-powered EdTech innovation for long term educational impact.

**Implications for K12 skills in the GenAI Era**

To anticipate the K12 skill shift, we need to consider: 1) evidence or cases on how GenAI has been impacting adult professionals; 2) what critical human development is required for K12 learners in comparison to adults. "Research by economists EdFelten, Manav Raj, and Rob Seamans concluded that AI overlaps with the most highly compensated, highly creative, and highly educated jobs. Only 36 job categories out of 1016 had no overlap with AI. Those jobs include dancers, athletes, roofers, etc., which are highly physical jobs. Ethan Mollick and other researchers conducted experiments on how GenAI impacts BCG consultant tasks (i.e. creative, analytical, writing & marketing, persuasiveness tasks), and the AI-powered consultants were faster, and their work was more creative, better written, and more analytical" (Mollick, E. 2024). GenAI is impacting how we work, 100%! Yet, adult professionals work as experts who direct GenAI tools with clear goals using various skills to refine AI output. K12 learners still need to acquire those skills to work with AI, but we can anticipate K12 skill shifts due to the impact on adult tasks.

Foundational cognitive skills (e.g. literacy and numeracy) are still critical for K12 students as these are gateway skills required to work with other skill sets. A child will not be able to critically analyze whether a piece of writing refinement advice from GenAI is useful or not, without being able to read. However, formal schooling in global K12 education systems need to 1) incorporate explicit cultivation of higher order thinking cognitive skills (e.g. critical thinking, creative thinking, problem solving) which are non-routine analytical skills hard to be automated by AI, metacognitive skills (e.g. self-regulated learning) that exhibits high quality thinking processes for children to effectively use learning resources and tools, especially GenAI tools; 2) pilot skills that are coined for the AI Age (e.g. AI literacy, computational and scientific inquiry, fusion skills); 3) explore the use of GenAI tools for creative thinking and global competence that enables learners to contribute to real world scientific and social problem solving; 4) encourage explicit development of unique human intelligence, for instance, social emotional skills (e.g. empathy, communication), physical (e.g. play sports, musical instruments, and making art and crafts) and practical skills (e.g. self-care, cooking). The value of living a healthy, happy, meaningful life will become more and more of a central topic in the GenAI age. This is also what children should care about, in addition to finding a place in the labor market. (Note. All skills mentioned above are listed in Table 1 in the appendix, which provides an updated K12 skills summary with definitions, sub-skills or skill specifics.)

**Implications for Evaluation of Learning in the GenAI Era**

The most researched commercial K12 EdTech products are predominantly intelligent tutoring systems, recommender systems and adaptive assessments that focus on foundational cognitive skill development in domain K12 subjects, because cognitive skills have been considered important determinants for success in the labor market. This is how learning has been assessed in high-stakes evaluations in global K12 education systems till now. However, this has to change, quickly! Evaluation works as a north star that guides education reform directions and EdTech innovation from the private sector. Simple truth, EdTech companies hardly invest resources in products that are not aligned with what Ministries of Education have been evaluating. Therefore, evaluation policies have to adapt to the change to encourage innovation. If K12 skills are to shift to critical thinking, creative thinking, problem solving, self-regulated learning, media and AI literacy, computational and scientific inquiry, global competence, education systems need to encourage private sector to innovate to support the instruction and evaluation of these skills, particularly solutions coming from GenAI-powered EdTech sector. Would standardized testing be enough to capture the development of these skills? No!

Large-scale K12 skill evaluation needs to expand from standardized assessment to allow tangible learner work portfolio (e.g. an app, a presentation on climate change, a creative visual design of future bicycles, etc.), that follows a constructivist approach. If the goal of learning is to prepare students to perform tasks from the changing labor market, then why not create that possibility already? If professionals now use GenAI to transform work quality and productivity, why do education systems still evaluate students solely through traditional cognitive test scores? A segment of K12 student leaders should be encouraged to pioneer this revolution, using GenAI tools to produce and contribute to society just like adult professionals. Technology is accessible to All, including K12 students; it's better we provide the right incentives to use.

**Strategic GenAI-powered EdTech Innovation for Long Term Educational Impact**

Table 2 in the attachment shows key stages of how technology has been incorporated into K12 education. Technology (including AI) has been shifting from digitizing learning inputs (e.g. textbooks and assessments), to intelligent tutoring systems that directly interact with learners at scale, to the GenAI era with individual learners being able to co-work with multimodal AI agent for iterative, collaboration tasks that lead to tangible outcomes (e.g. an app, a marketing pitch, creative design of future bicycle, etc.). In the GenAI era, AI agents can be a versatile educational partner - serving as a multimodal tutor, an analytical collaborator, case study mentor, role-play counterpart, creative brainstorm guide and project assistant. The educational possibilities are limitless.

Below are market gaps identified based on Table 1 updated summary of K12 skills and Table 2 K12 Education & EdTech Evolution: 1) existing K12 EdTech skill development products predominantly focus on cognitive skill acquisition and evaluation, in the form of tutoring systems and adaptive assessments; 2) there is almost no commercial product that explicitly focus on the cultivation and evaluation of critical thinking, creative thinking, problem solving, self-regulated learning, media and AI literacy, computational and scientific inquiry, global competence. There are small scale pilots for certain skills, but priority is not given on a system level required change, and the EdTech private sector is best positioned to innovate.

Below are a few strategic recommendations for EdTech sector to adopt GenAI,

1. GenAI-powered EdTech products that explicitly develop or evaluate the following skills,
    - Higher order thinking cognitive skills: critical thinking, creative thinking, problem solving;
    - Metacognitive skills: self-regulated learning
    - Skills coined for the AI age: media and AI literacy, computational and scientific inquiry;
    - Skills enabling learners to solve real world problems: creative thinking, global competence.

2. GenAI enhancement of existing ITS for foundational cognitive skill development and allowing poor-resourced educational contexts to use enhanced ITS with minimum technical and infrastructural requirements. For instance, multimodal AI agent capabilities could allow real time, dynamic, nuanced natural language interaction between individual learner and ITS system, improving learner agency, self-regulated learning and real-time individualized feedback and conceptual correction.

3. Serious R&D is required to design and develop the above solutions to avoid de-skilling K12 learners of high quality thinking and problem solving journey they have to go through. For example, when intelligent tutoring systems scaffold difficult math world problems by reducing the difficulty level of instructional language, or by breaking a complex problem down to simple steps for learners to solve, this is taking away some of the problem solving journeys from the learner. The question is, to whom should ITS provide such support, when and how ITS should provide such assistance. For creative thinking tasks, GenAI tools can be used to generate ideas, generate multiple ideas or help evaluate ideas very quickly. For instance, in 2022 PISA creative thinking framework, 15 year olds are asked to "think of 3 original improvements that can be made to a standard bicycle, clearly explain how each idea works within 5 min" (OECD, 2019). A critical issue to be tackled is, how to design learner-AI interactions in scenarios where learners co-work with AI agents for outputs. Learners command, project manage, iterate and evaluate AI output instead of relying on copying AI answers. Who has the ownership of the output when learners and AI both contributed to the work? What essential product mechanisms are needed to enhance human intelligence and keep learners accountable? These require serious R&D from interdisciplinary expertise and through effective collaboration, quick experimentation and pilot testing.

# Appendix

Table 1. Global K12 Skills Summary and Update

| Skill | Definition | Sub-skills or Skill Specifics |
|---|---|---|
| **Cognitive skills** | "A set of thinking strategies that enable the use of language, numbers, reasoning and acquired knowledge. They comprise verbal, nonverbal and higher-order thinking skills" (OECD a, 2019). | 1. Foundational literacy & numeracy skills assessed in K12 domain subjects (e.g. English, Math, Science, Arabic, etc.)<br><br>2. Higher order thinking skills (e.g. critical thinking, creative thinking, problem solving.) |
| **Metacognitive skills** | "Learning-to-learn skills and the ability to recognise one's knowledge, skills, attitudes and values. Non-routine analytical skills, where awareness of one's own learning and thought process leads to the intentional application of specific learning techniques to different situations"(OECD a, 2019). | Example: self-regulated learning |
| **Social emotional skills** | "A set of skills manifested in consistent patterns of thoughts, feelings and behaviors that enable people to develop themselves, cultivate relationships at home, school, work and in the community, and exercise their civic responsibilities"(OECD a, 2019). | Big five framework<br>1. Collaboration: empathy, trust, cooperation<br>2. Open-mindedness: curiosity, creativity, tolerance<br>3. Emotional regulation: stress resistance, optimism, emotional control<br>4. Task performance: self-control, responsibility, persistence<br>5. Engaging with others: sociability, assertiveness, energy<br>Achievement motivation & Self-efficacy |
| **Physical skills** | "A set of abilities to use physical tools, operations and functions, including manual skills to use machines and ICT devices; and ability to mobilize one's capabilities including strength, muscular flexibility and stamina"(OECD a., 2019). | Example: play sports, musical instruments, craft artworks |
| **Practical skills** | "Skills that are required to use and manipulate materials, tools, equipment and artefacts to achieve particular outcomes. Practical skills are often associated with manual dexterity and craftwork". (OECD a, 2019). | Example: cooking, engaging in written work or using various technologies. |
| **Learning in the digital world** | "The capacity to engage in an iterative and self-regulated process of knowledge building and problem solving using computational tools and practices" (OECD, 2023). | 1. Self-regulated digital learning: monitoring and control of one's metacognitive, cognitive, behavioral, motivation and affective processes while learning.<br><br>2. Computational and scientific inquiry practices: capacity to use digital tools to explore systems, represent ideas, solve problems with computational logic. |
| **AI literacy** | "Skills that equip students to build and work with AI to meet the needs of a shifting workforce, which include teaching students what AI is, enabling them to create with it, think critically about its impact, and advocate for responsible use" (Williams, R. et al., | 1. Key outcomes:<br>- <u>Technical AI knowledge</u>;<br>- <u>Ability to think critically about the implications of AI</u>;<br>- <u>Ability to apply AI knowledge</u>.<br>2. Pedagogical design strategies: |
| | 2022). | - <u>Active learning</u> (by engaging in hands-on activities, process information through reflection);<br>- <u>Embedded ethics</u> (by teaching technical and ethical concepts in tandem; enabling children to view technology as a sociotechnical system and critique the ethical implications of specific technologies);<br>- <u>Low barriers to access</u> (by centering student and teacher needs in the design with the goal to reach all students with AI education, and incorporating AI with subjects like art, dancing, robotics, etc). |
| **Creative Thinking** | "The ability to generate, evaluate and improve ideas to produce original and effective solutions, advance knowledge and create impactful expressions of imagination, and to tackle emerging challenges creatively" (OECD b, 2019). | Three core skills:<br>1. Generate diverse ideas;<br>2. Generate creative ideas;<br>3. Evaluate and improve ideas.<br>Four main domains of core skill application:<br>1. Written expression;<br>2. Visual expression;<br>3. Scientific problem solving;<br>4. Social problem solving. |
| **Global Competence** | " A multi-dimensional construct that requires a combination of knowledge, skills, attitudes and values successfully applied to global issues or intercultural situations." (OECD, 2018). | Four dimensions of global competence:<br>1. Examine local, global and intercultural issues;<br>2. Understand and appreciate the perspectives and world views of others;<br>3. engage in open, appropriate and effective interactions across cultures;<br>4. Take action for collective well-being and sustainable development.<br>Six levels of proficiency are defined in the global competence assessment framework. |
| **Fusion skills** | "The combination of creative, entrepreneurial and technical skills that enable workers to shift to new occupations as they emerge" (OECD a, 2019). | Life long learning agility (Author expanded)<br>- Continuous learning mindset<br>- Ability to update knowledge and skills<br>- Career navigation<br>- Creative technical integration (e.g. human-AI collaboration) |
| **2029 Media and AI literacy** | The assessment framework is still being defined by OECD PISA still, so there's no concrete definition yet. However, media and AI literacy is a new set of skills in response to the world that is mediated by digital and AI tools. It is considered a "set of competencies to interact with digital content and platforms effectively, ethically and responsibly. The new assessment is envisioned as a simulated environment that would allow the collection of evidence for multiple competencies of the literacy model. These competencies are assessed using a variety of functional tools that are accessible to students in a realistic way through the assessment (e.g. realistic simulations of the internet, social media, and generative AI tools)"  (OECD, 2025). | |

Table 2. K12 Education & EdTech Evolution

| | Traditional Classroom Teaching & Learning | Technology Enhancement of Learning Inputs | E-learning and AI-powered Adaptive Learning | GenAI Era |
|---|---|---|---|---|
| Key characteristic | Human-to-human, face-to-face instruction, physical textbook & | Use of computers, digital white boards, and digital curriculum as learning inputs | LMS (learning management system), open online learning resources (e.g. MOOC, youtube), AI-powered | Multimodal content and curriculum dynamic customization;<br>Multimodal GenAI enhanced ITS systems;<br>GenAI agents enabling learner-AI agent |
| | assessments | | adaptive learning systems (e.g. ITS intelligent tutoring systems) | coworking and iterative collaboration; |
| Key interaction for skill development | Human to human, no involvement of technology. Lack of personalization. | Human to human, technology acts only as supporting inputs. Lack of personalization. | Learner-AI interaction, with human teachers in the loop. Personalization is done to a limited scope, not fully on an individual level. | Multimodal GenAI Agents directly interact with individual students for skill development; Human-AI interaction is more natural, dynamic and nuanced than that with traditional AI system; Personalization fully executed on an individual level. |
| Involvement of technology and impact evaluation of technology (AI & GenAI) | NA | Minimum involvement, indirectly, only as learning inputs. Impact is assessed through standardized test scores that measure foundational cognitive skills. | Directly involved in the skills development process; ITS is the most researched learning system and commercial products; Impacted is assessed through standardized test scores that measure foundational cognitive skills. Lack of evidence on higher order cognitive thinking skills (e.g. critical thinking, creative thinking, problem solving), metacognitive skills. | Directly and profoundly; More creative designs the leverage GenAI agent for various skills (e.g. creative thinking, critical thinking) would rise; Impact evaluation will expand to incorporate learner work portfolios as output working with GenAI systems, in addition to standardized assessments. |
| Skills and Evaluation | Cognitive skills considered important determinants of success in the labor market; Evaluation: Standardized assessment of foundational cognitive skills; | Cognitive skills considered important determinants of success in the labor market; Evaluation: Standardized assessment of foundational cognitive skills. | Cognitive skills considered important determinants of success in the labor market; Social emotional skills also considered key predictors of occupational success; Metacognitive skills (e.g. self-regulated learning) has been found to be a strong predictor of effective use of adaptive learning systems; Non-routine higher order thinking cognitive skills (e.g. critical thinking, creative thinking) considered not able to be automated by AI; Evaluation: standardized digital adaptive assessments | Higher-order thinking cognitive skill (e.g. critical thinking, creative thinking, problem solving), metacognitive skills (e.g. SRL), AI age skills (e.g. media and AI literacy), skills to focus on creatively solving real world problems (e.g. creative thinking, global competence) are believed to be demanded by the changing labor market disrupted by GenAI; Evaluation: OECD has released assessment frameworks for the above skills, and some of the skills have been tested in pilot education systems already. Individual countries and education systems will still assess foundational cognitive skills, but slowly incorporate new skill evaluation through tangible learner work portfolio. K12 learning and tasks will move closer to changing the labor market, through simulated |

| | | | of foundational cognitive skills. | learning experience and evaluation. |
|---|---|---|---|---|
| Learner agency and human intelligence | Teacher-directed teaching and learning; Student as recipient of prescriptive knowledge and information. | Teacher-directed teaching and learning; Student as recipient of prescriptive knowledge and information. | Teacher-directed teaching and learning (mainstream formal schooling), co-exist with student-centered adaptive learning; ITS system design is student-centered, but pedagogical approach is prescriptive in nature ; Critical learner agencies (e.g. self-system factors, self-regulated learning, metacognitive skills) are needed to effectively utilize ITS to benefit learning gains. | Critical learner agency could be enhanced and explicitly cultivated through designs using GenAI agents (such as SRL, creative thinking, critical thinking); Learner co-work with AI agents for various skill development will rise; More emphasis on creative application of skills in solving real world challenges; Uniquely human intelligence: social emotional skills, practical and physical skills will be valued more than before. |